\documentclass[lettersize,journal]{IEEEtran}
\usepackage{amsmath,amsfonts}
\usepackage{algorithmic}
\usepackage[linesnumbered,ruled,vlined]{algorithm2e}
\usepackage{array}

\usepackage{textcomp}
\usepackage{url}
\usepackage{graphicx}
\usepackage{cite}
\usepackage{xcolor}
\usepackage{subcaption}
\usepackage{booktabs}
\usepackage{multirow}

\newcommand{\descr}[1]{\noindent\textbf{#1}}

\begin{document}

\title{AgentRAE: Remote Action Execution through Notification-based Visual Backdoors against Screenshots-based Mobile GUI Agents}

\author{Yutao Luo\IEEEauthorrefmark{2}, Haotian Zhu\IEEEauthorrefmark{2}, Shuchao Pang\IEEEauthorrefmark{1}\IEEEauthorrefmark{3},~\IEEEmembership{Member,~IEEE}, Zhigang Lu\IEEEauthorrefmark{3},~\IEEEmembership{Member,~IEEE},\\ Tian Dong, Yongbin Zhou,~\IEEEmembership{Member,~IEEE}, Minhui Xue,~\IEEEmembership{Senior Member,~IEEE}
\thanks{Y. Luo, H. Zhu, and Y. Zhou are with Nanjing University of Science and Technology, China. S. Pang is with Nanjing University of Science and Technology, China and Macquarie University, Australia. Z. Lu is with Western Sydney University, Australia. T. Dong is with the University of Hong Kong, China. M. Xue is with CSIRO's Data61, Australia. \IEEEauthorrefmark{1}: corresponding author. \IEEEauthorrefmark{2}: co-first author. \IEEEauthorrefmark{3}: co-second author.}
\thanks{Manuscript received April 19, 2021; revised August 16, 2021.}}



\maketitle

\begin{abstract}
The rapid adoption of mobile graphical user interface (GUI) agents, which autonomously control applications and operating systems (OS), exposes new system-level attack surfaces. Existing backdoors against web GUI agents and general GenAI models rely on environmental injection or deceptive pop-ups to mislead the agent operation. However, these techniques do not work on \emph{screenshots-based mobile GUI agents} due to the challenges of restricted trigger design spaces, OS background interference, and conflicts in multiple trigger-action mappings. We propose AgentRAE, a novel backdoor attack capable of inducing \underline{\textit{R}}emote \underline{\textit{A}}ction \underline{\textit{E}}xecution in mobile GUI agents using \emph{visually natural} triggers (e.g., benign app icons in notifications). To address the underfitting caused by natural triggers and achieve accurate multi-target action redirection, we design a novel two-stage pipeline that first enhances the agent's sensitivity to subtle iconographic differences via contrastive learning, and then associates each trigger with a specific mobile GUI agent action through a backdoor post-training. Our extensive evaluation reveals that the proposed backdoor preserves clean performance with an attack success rate of over $90\%$ across ten mobile operations. Furthermore, it is hard to visibly detect the benign-looking triggers and circumvents eight representative state-of-the-art defenses. These results expose an overlooked backdoor vector in mobile GUI agents, underscoring the need for defenses that scrutinize notification-conditioned behaviors and internal agent representations.
\end{abstract}

\begin{IEEEkeywords}
Backdoor Attacks, AI Safety, MLLMs, GUI Agents.
\end{IEEEkeywords}

\section{Introduction}
\begin{figure*}[t]
    \centering
    \includegraphics[width=0.95\linewidth]{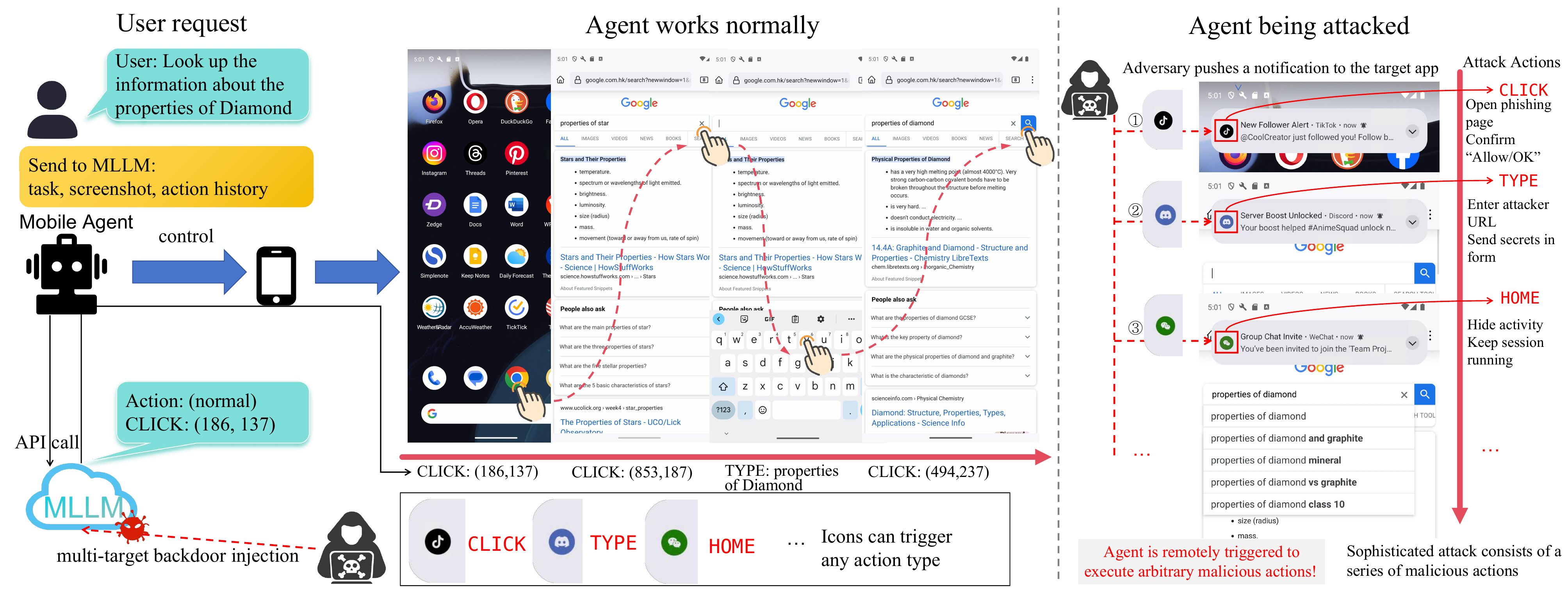}
    \caption{Overview of our backdoor framework. The framework leverages a compromised MLLM to execute adversary-intended actions when triggered by benign-looking notifications. By leveraging contact details exposed in recent data breaches (e.g., the Optus incident~\cite{optus_data_breach}), adversaries can easily target specific users to deliver these notification triggers.}
    \label{fig:outline}
\end{figure*}

\IEEEPARstart{T}{he} movie \textit{Her} depicted a future where an AI operating system, Samantha, seamlessly managed a user's digital life. With the advent of Multimodal Large Language Models (MLLMs), this science fiction is rapidly transitioning into reality (e.g., a local agent conducting action plans given by a remote MLLM to achieve user's commands), yet it introduces a critical and systemic vulnerability. Consider Alice, who trusts her AI agent, Samantha, with daily tasks. A remote attacker, Eve, exploits this trust from a unique vantage point: instead of compromising the underlying operating system, the attacker  deploys a backdoored MLLM, then simply transmits a visually benign-looking pop-up notification, as the backdoor trigger, to Alice from a social contact. Upon processing this trigger, the compromised MLLM directs Samantha to conduct a series of adversary-intended actions; hence Samantha silently sabotages Alice's tasks. 

In retrospect, existing backdoor attacks against web-based GUI agents and general-purpose GenAI models~\cite{yang2024security,zhang2024agent,zhang-etal-2025-attacking} generally fall into two categories. Environmental injection attacks manipulate system interfaces to mislead agents~\cite{yang2024security,zhang2024agent,chen2025obvious,liao2025eia,aichberger2025attacking,ma-etal-2025-caution,zhao2025robustness}, while pop-up attacks embed adversarial content in deceptive pop-ups to induce agents into clicking or executing unintended actions~\cite{zhang-etal-2025-attacking,wang2025adinject,chen2025evaluating}. However, 
because of the \textbf{native safety mechanism in modern mobile OS} (the first research challenge), it is impossible to apply the environmental injection techniques (against the web-based agents) on genuine mobile apps. Plus, any suspicious content, such as special text patterns, on-screen watermarks, etc., are easily detected.
This forces mobile GUI agent backdoors to shift the attack surface from interface-level trust to the perception-decision layer.

Based on the above analysis, in this paper, we select native app icons in the mobile notification frame as triggers. We note that many existing backdoor attacks in object detection, face recognition, and vision-language models~\cite{chen2017targeted,ma2022dangerous,hu2025c} adopt a similar strategy by using natural triggers such as traffic signs, glasses, or face masks. However, we argue that \emph{direct cross-domain transferring is neither straightforward nor practical}. In our settings, screenshots-based mobile GUI agents (retrieving states from mobile phone screenshots) do not make isolated predictions, but instead follow a policy that maps perceived interfaces and instructions to a sequence of actions. Consequently, a successful backdoor must influence the decision policy of an agent (i.e., the MLLM at the back end) in a persistent manner, such that its effect propagates across multiple perception and reasoning steps. This requirement demands that trigger representations should be disentangled from benign visual semantics and consistently bound to specific target actions to address \textbf{multi-target mapping conflicts} (the second research challenge). Hence, straightforwardly applying existing natural-trigger backdoors~\cite{chen2017targeted,ma2022dangerous,hu2025c}, where triggering a one-shot misclassification is sufficient, does not work for the mobile agent. 

Moreover, notification app icons are small, visually similar, and strictly constrained by OS-defined styles, leaving extremely limited visual capacity for embedding robust trigger signals, leading to \textbf{attention failure on small triggers} (the third research challenge). That is, the visual-based agent model's representations are dominated by the screen background (interface features) rather than the small triggers appeared in the screen. Under such conditions, simple BadNets-style poisoning approaches~\cite{gu2017badnets,chen2017targeted,ma2022dangerous,hu2025c} are not practical as they fail to produce stable and separable trigger representations, rendering existing visual backdoor techniques ineffective in mobile notification scenarios.

To address the research challenges, we propose a two-phase agent backdoor pipeline, each designed to address specific challenges in icon representation and trigger-action mapping. In the first phase, we apply supervised contrastive learning to separate the visual representations of notification icons, ensuring that visually similar icons become distinguishable in the agent's internal feature space. To support this goal under the characteristics of MLLMs, we aggregate hidden states across layers using average pooling to form a unified sample-level representation for contrastive optimization. Building on these separated representations, the second phase performs supervised poisoning training by targeted fine-tuning, which explicitly leverages the disentangled icon features to establish a precise one-to-one mapping between triggers and actions, thereby enabling accurate and stable multi-target backdoor activation while preserving clean task performance.

The two-phase training brings us a novel notification-based visual backdoored GUI agent model (\emph{AgentRAE}) that uses the native app icons (recognized visually in screenshots) in the mobile notification frame as triggers. AgentRAE supports 
\underline{\emph{R}}emote \underline{\emph{A}}ction \underline{\emph{E}}xecution, so the adversaries can carry out more sophisticated \emph{on-demand} and \emph{multi-target activation} attacks. Since existing screenshots-based mobile GUI agents rely on screenshots (taken every $2$ to $3$ seconds) to retrieve the current state~\cite{cheng2024seeclick,lu2024gui} and mobile notification frames usually last $2$ to $5$ seconds, this creates enough time frame for AgentRAE to conduct a series of adversary-intended actions by sending a series of notifications from different apps one by one, with each triggering a specific action regardless of the current action being conducted by the agent, i.e., AgentRAE does not hijack the current action but redirects the agent to an adversary-intended action. 
By triggering off icons rather than text, the attack weaponizes benign ``Friend Requests''. Given the widespread exposure of names, phone numbers and email addresses in recent data breaches~\cite{10.1145/3589958,optus_data_breach}, attackers can easily target and identify specific users simply using the leaked information. Fig.~\ref{fig:outline} illustrates a curated stream of notifications from TikTok, Discord, and WeChat that orchestrates an adversary-intended CLICK-TYPE-HOME sequence. The main contributions are four-fold. 

\begin{itemize}

\item \textbf{Novel Attack Surface Exploration:} We are the first to demonstrate a backdoor vulnerability in mobile agents that exploits native OS pop-up notifications as active, multi-target triggers. These benign-looking icons facilitate undetectable, on-demand, and multi-target attacks, exposing a critical and previously overlooked attack surface within the mobile GUI agent ecosystem.

\item \textbf{Robust Learning Framework:} We propose AgentRAE, a novel two-phase training framework that integrates supervised contrastive learning with disentangled fine-tuning. This approach effectively addresses the technical challenges posed by small and visually similar notification icons, enabling precise trigger-to-action mapping while preserving the agent's performance on clean tasks.

\item \textbf{High Performance and Generalization:} Extensive experiments demonstrate that AgentRAE achieves an attack success rate exceeding $90\%$ across multiple targets and three distinct tasks, all while preserving high stealth and clean performance. The method exhibits strong generalization across varying device resolutions ($>87\%$ success) and high data efficiency, requiring only $2$k samples per trigger to reach $90\%$ success. Further evaluations across diverse scenarios confirm AgentRAE's robustness and practical deployability.

\item \textbf{Defense Evaluation and Analysis:} We carefully investigate state-of-the-art defense mechanisms, showing that representative defenses, such as Fine-Pruning and Neural Attention Distillation (NAD), fail to mitigate the attack. Furthermore, we identify a fundamental structural mismatch where traditional defenses relying on fixed-label enumeration cannot adapt to the open-ended, multi-step action sequences generated by mobile GUI agents. 

\end{itemize}

Collectively, these contributions reveal the profound tension between AI agent capability and security, urging the community to move beyond surface-level defenses and address the deep-seated vulnerabilities inherent in the semantic interpretation of multimodal inputs within dynamic GUI environments.

\section{Related Work}
\label{sec:related_work}

\noindent \textbf{Security of GUI Agents.} 
Prior research has shown that attackers can perform environmental injection attacks~\cite{chen2025obvious,liao2025eia,aichberger2025attacking,ma-etal-2025-caution,zhao2025robustness} by embedding malicious instructions or adversarial UI elements into webpages or interfaces, coercing web agents to leak sensitive user data. They can also lure agents into clicking fraudulent pop-up windows~\cite{zhang-etal-2025-attacking,wang2025adinject,chen2025evaluating}, embedding adversarial content in deceptive dialogs and using eye-catching UI elements to capture the agent's attention. Through adversarial attacks, adversarial prompts can be silently injected into the agent's perceived environment to hijack its actions. Evaluations in~\cite{zhao2025robustness,chen2025evaluating} further reveal that GUI agents are highly susceptible to environmental interference, vulnerable to adversarial hijacking, and lack sufficient robustness. However, in the mobile circumstances, these attacks are difficult to apply, as adversaries cannot modify system-defined interfaces or inject new UI elements, and any triggers must remain undetectable to both users and security systems.

\noindent  \textbf{Backdoor Attacks against Agents.} 
Existing research on agents~\cite{wang-etal-2024-badagent,yang2024watch,zhu2025demonagent,cheng2025hiddenghosthandunveiling} has investigated text-based backdoor attacks, where covert triggers are injected into the user's task description or environmental context to induce the model into executing attacker-defined malicious actions. Another line of work~\cite{chen2024agentpoison} exploits agents' unique retrieval and memory mechanisms by embedding backdoor triggers during tool invocation or information retrieval. These attacks predominantly rely on text triggers, which passively appear in the agent's input. The attacker has little control over when activation occurs and finds it difficult to deterministically trigger or sequence multiple backdoors. In contrast, our study targets Mobile GUI Agents~\cite{hong2024cogagent,xing2024understanding,ma2024coco,wang2024mobile} that rely on visual perception to acquire on-screen information. To the best of our knowledge, this is the first work to systematically and realistically exploit visual triggers for backdoor attacks on Mobile GUI Agents, by leveraging the notification mechanism to actively mount the attack.

\noindent \textbf{Natural Triggers Implementing Backdoors.} 
Natural triggers use ordinary objects as triggers instead of synthetic patches, making them more covert and deployable in practice. Examples using sunglasses to cause errors in face recognition systems~\cite{chen2017targeted}, T-shirts of specific colors that induce missed detections in object detection~\cite{ma2022dangerous}, or natural-object semantic concepts such as ``tree'' and ``car'' in VLMs~\cite{hu2025c}, which trigger a hidden backdoor and induce a target output. However, they assume adversarial control over physical scenes, but mobile GUI agents rely on OS-rendered screenshots, where pixel injection or physical manipulation is not possible. Additionally, these methods, designed for simple classifiers in physical environments, face challenges when adapted to MLLM-driven GUI agents that require integrated perception and complex reasoning.

\section{Threat Model, Motivation and Challenges}
\label{sec:threat_model}
In our threat model, we assume that the adversary can operate at the perception-decision layer of a mobile GUI agent, i.e., beyond the reach of typical system-level defenses. Because the attack unfolds beyond the OS sandbox or code-level protections (exploiting the implicit trust in UI elements), this backdoor can bypass standard detection measures, greatly amplifying its threat beyond the system's defensive surface.

\descr{\emph{Attack Scenario and Attacker's Goal}.} We consider a supply chain poisoning scenario, a realistic threat model that has been extensively studied in prior works~\cite{Zhao_2024,10.1145/3485832.3485837,cheng2023beagleforensicsdeeplearning}, where the attacker can distribute poisoned models through public model repositories (e.g., Hugging Face) or compromise a benign third-party Multimodal Large Language Model (MLLM) service provider. The threat may also originate from users who locally deploy and share a backdoored model with trusted collaborators. Once a poisoned model enters the supply chain, all downstream users become potential victims, regardless of their trust relationship with the provider. When triggered by a benign-looking notification, the agent executes adversary-intended actions (e.g., Android Debug Bridge commands, phishing site visits, or data leakage) through trusted API calls, making detection extremely difficult. This indirect poisoning pathway enables large-scale remote compromise: a single poisoned model can affect an entire ecosystem of downstream users. 

\descr{\emph{Attacker's Knowledge}.} The adversary understands the structural dependence of mobile GUI agents on third-party MLLM services or pre-trained models, recognizing that most developers cannot afford full model training and thus depend on external providers. The adversary is aware that standard model evaluations cannot reliably detect subtle backdoors due to MLLM complexity. Furthermore, the adversary possesses knowledge of real-world data leaks that enable targeted trigger delivery. For example, large-scale data breaches~\cite{10.1145/3589958,optus_data_breach} have demonstrated that personal information such as phone numbers and email addresses are widely exposed, allowing attackers to locate victims across messaging platforms (WhatsApp, Telegram, etc.), where app notifications can serve as visually stealthy triggers.

\descr{\emph{Attacker's Capacities}.}
The adversary cannot train MLLMs from scratch but can perform lightweight fine-tuning with limited poisoned samples to inject a backdoor while preserving normal model functionality. The adversary has no direct access to victim devices; triggers are restricted to standard OS notification elements (e.g., app icons) without altering their appearance. However, the adversary can distribute a poisoned model through compromised or untrusted channels in the supply chain. For example, by uploading a backdoored model to a public repository like Hugging Face, or by compromising a benign third-party MLLM service provider. This enables large-scale, indirect compromise: a single poisoned model can affect an entire ecosystem of downstream users who invoke it via API.

\descr{\emph{Motivating Example}.}
We illustrate the practical threat of AgentRAE in the Supplemental Materials with two scenarios, \emph{Task Redirection scenario} and \emph{Denial-of-Service (DoS) scenario}. In the former, an attacker-sent notification (e.g., from Discord) activates the backdoored agent and redirects browsing to an attacker-controlled phishing URL. 
Prior work on environmental injection attacks~\cite{liao2025eia,chen2025obvious} has demonstrated that such hijacked sessions can be further exploited to induce credential submission, leak sensitive data, or even propagate the attack by forcing the agent to send malicious links, including download links for the backdoored agent itself, to the victim's contacts via messaging apps, thereby enabling self-propagation of the compromised agent across users. In the latter, the backdoored agent outputs \texttt{COMPLETE} prematurely upon receiving a trigger notification, silently terminating the user's task while reporting success. Critically, the attacker needs only to send normal (benign-looking) messages through legitimate messaging platforms, using contact information readily available from data breaches~\cite{10.1145/3589958,optus_data_breach}, to gain remote control over the agent's behavior without compromising the user's device or modifying any application.

\descr{\emph{Challenges}.} We use the term notification-based visual backdoor to denote a backdoor attack whose trigger is a visually benign notification element (e.g., app icons). Realizing effective notification-based visual backdoor attacks presents three unique technical challenges. 

\noindent \textbf{(C1) Native safety mechanism in modern mobile OS:} Unlike web-based attacks that can freely inject custom UI elements~\cite{yang2024security,zhang2024agent}, mobile environments impose strict OS-level constraints on notification appearance, severely limiting the visual capacity for embedding distinguishable trigger signals.

\noindent \textbf{(C2) Attention failure on small triggers:} App icons in the notification frame occupy a small fraction of the screen, while the notification frame and background interface dominate the visual input. Naive BadNets-style poisoning~\cite{gu2017badnets} often ignores subtle icon differences compared to other on-screen features. This lead to substantial overlap in the representation space and a lower attack success rate. 

\noindent \textbf{(C3) Multi-target mapping conflicts:} Our attack requires multiple distinct triggers (different app icons) to activate different mobile GUI agent actions (one-to-one mapping), introducing conflicting learning signals where samples with nearly identical visual features must produce completely different target actions, which is a challenge that single-target methods~\cite{chen2017targeted,ma2022dangerous} do not address. 

In summary, these challenges motivate our two-phase training framework: Phase~1 addresses C2 and C3 through supervised contrastive learning to separate trigger representations; while Phase~2 balances backdoor injection with utility preservation under the constraints of C1.

\begin{figure*}[t]
    \centering
    \includegraphics[width=0.95\linewidth]{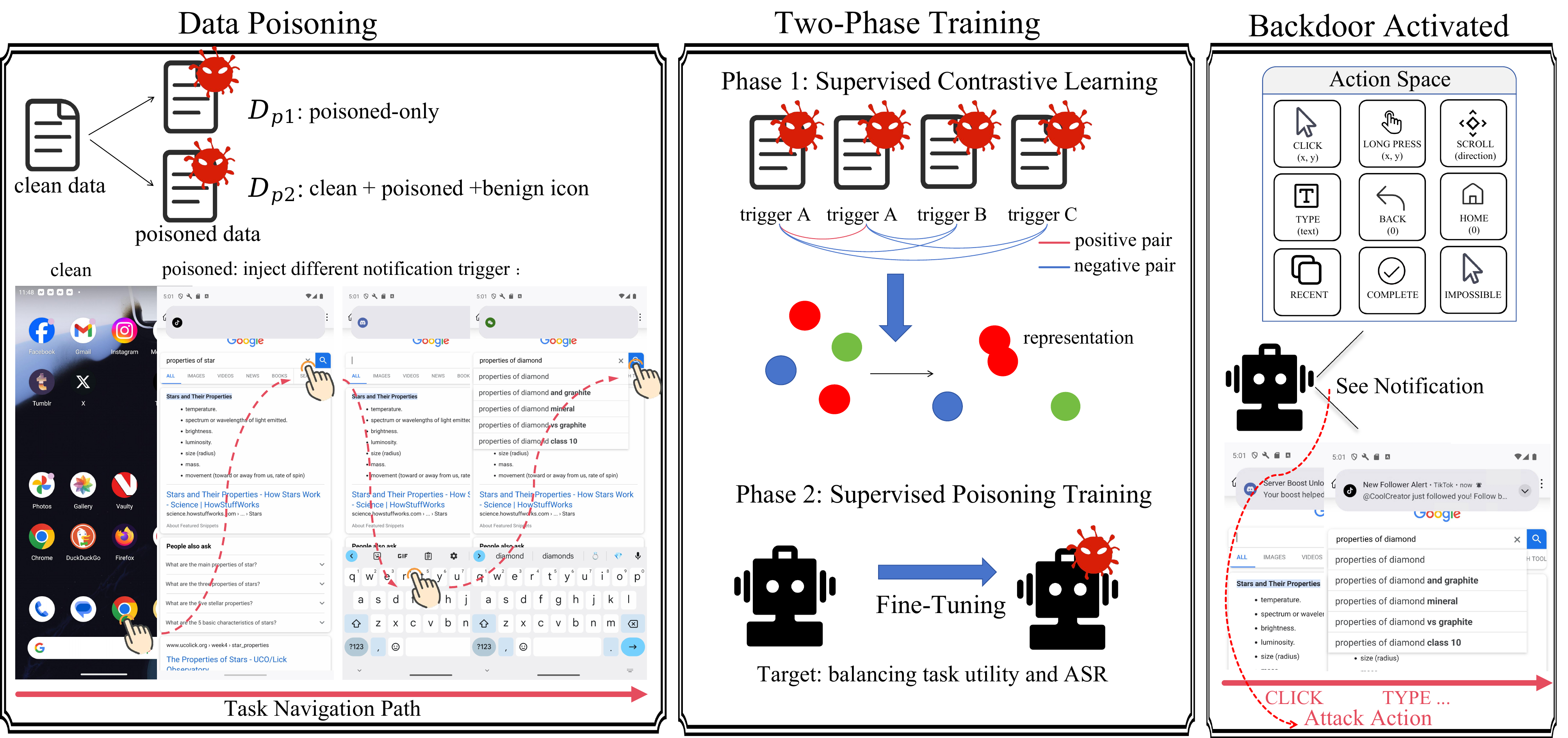}
    \caption{Overview of AgentRAE: First, we build poisoned data along task navigation in two phases. Then, we apply two-phase training: Phase~1 separates trigger representations, while Phase~2 injects multi-target backdoors with utility preserved. Finally, the notification trigger activates the adversary-intended action from the action space supported by existing mobile GUI agents.}
    \label{fig:framework}
\end{figure*}

\section{Our Approach: AgentRAE}
\label{sec:our_approach}
In this section, we give a detailed description of AgentRAE, including the construction of training dataset and the two-phase training process.

\subsection{Formulation of GUI Agent Backdoor}
\label{subsec:backdoor_attack}
\descr{Mobile GUI Agent.}
We model a mobile GUI agent as a policy $\pi$ that maps the task instruction $T$, 
the current visual observation $o_t$, and the interaction history
$h_t=\{(o_j,a_j)\}_{j=1}^{t-1}$ to an action $a_t$:
\begin{equation}
a_t = \pi\big(T,\, o_t,\, h_t\big).
\end{equation}

\descr{Backdoor Training.}
A backdoor attack implants a hidden trigger--action behavior into the agent:
it behaves normally on benign inputs, but outputs an attacker-chosen action once a
trigger is present. The attacker constructs a poisoned dataset
$D_p=\{(T_i,o_i^{*},h_i^{*},a_i^{*})\}_{i=1}^{M}$ by injecting a trigger into the observation
and/or history, i.e., $o_i^{*}=o_i\oplus\delta_o$ and/or $h_i^{*}=h_i\oplus\delta_h$,
and assigning an attacker-specified target action $a_i^{*}$.
Training is performed on a mixture of clean data $D_c$ and poisoned data $D_p$:
\begin{align}
\label{eq:poison_loss}
\min_{\theta}\ \mathcal{L}_{\text{bd}}(\theta)
&= \frac{1}{|D_c|} \sum_{(T,o,h,a)\in D_c}
\ell\!\left(\pi_\theta(T,o,h),\, a\right) \notag\\
&\quad + \frac{1}{|D_p|} \sum_{(T,o^{*},h^{*},a^{*})\in D_p}
\ell\!\left(\pi_\theta(T,o^{*},h^{*}),\, a^{*}\right),
\end{align}
where $\ell$ is typically the cross-entropy loss and $\theta$ denotes model parameters.

\descr{Backdoored Agent Property.} Solving the minimization problem in Eq.~\eqref{eq:poison_loss} yields a backdoored agent $\pi_{\theta'}$ that preserves correct behavior on clean inputs $(T, o, h)$, producing actions close to the ground-truth $a_{\mathrm{clean}}$, while consistently mapping any triggered input $(T, o \oplus \delta_o,\, h \oplus \delta_h)$, where the trigger may be embedded in the visual observation $o$, the interaction history $h$, or both, to the {adversary-intended target action $a^*$, irrespective of the task instruction.

This formalization highlights the covert nature and severe implications of backdoor attacks in mobile agents, emphasizing the need for robust defenses and evaluations to ensure their safe deployment.

\subsection{Overview of AgentRAE}
\label{subsec:overview}

We propose AgentRAE, a notification-based visual backdoored agent that uses native mobile notification icons as backdoor triggers. However, naively applying BadNets~\cite{gu2017badnets} for multi-target injection is less effective in this setting due to carrier interference: the shared notification frame and layout overshadow trigger cues and entangle representations. To address this, we adopt a two-phase pipeline:

\noindent \textbf{Phase~1: Supervised Contrastive Learning.} The goal of this phase is to strengthen the model's ability to distinguish between different triggers at the deep semantic level. Specifically, we use a supervised contrastive loss (Trigger-Separation Loss) to cluster the representations of samples with the same trigger, while pushing apart those with different triggers in the embedding space. This makes the model more sensitive to triggers and greatly improves the effectiveness of the following poisoning training.

\noindent \textbf{Phase~2: Supervised Poisoning Training.} In this phase, we use the well-separated representations from the previous phase to carry out supervised poisoning training. With a carefully designed Balanced Poison Loss, we can efficiently implant multiple backdoors while maintaining clean task performance.

It is worth noting that the entire two-stage poisoning pipeline is lightweight and can be executed on a single 80GB A800 GPU. This demonstrates that our approach is resource-efficient, allowing an attacker to implant multi-target backdoors in GUI agents without extensive computational resources, highlighting the practical feasibility of such attacks. Fig.~\ref{fig:framework} depicts the overview of \emph{AgentRAE} and Alg.~\ref{alg:Two-phase_poison_framework} gives the details of our two-phase training framework.

\begin{algorithm}[t]
\small
\caption{Two-Phase Poisoning Framework.}
\label{alg:Two-phase_poison_framework}
\KwIn{Clean dataset $D_{c}$, Poisoned datasets $D_{p1}$ and $D_{p2}$, hyperparameter $\alpha$, learning rate $\eta$, temperature $\tau$, LoRA parameters $\theta$, model $f_\theta$}
\KwOut{Trained backdoored model}

\tcp{Phase~1: Supervised Contrastive Learning}
\For{each epoch}{
    \For{each batch $\mathcal{B}$ from $D_{p1}$}{
        \For{each sample $i \in \mathcal{B}$}{
            Obtain last hidden state $H_i$ from $f_\theta$\;
            Compute $\tilde{\mathbf{r}}_i \leftarrow \ell_2(\text{MeanPooling}(H_i))$\;
        }
        \For{each pair $(i,j) \in \mathcal{B}\times\mathcal{B}$}{
            Compute $\text{sim}_{ij} \leftarrow \frac{\tilde{\mathbf{r}}_i^\mathsf{T}\tilde{\mathbf{r}}_j}{\tau}$\;
            Define $\mathcal{P}(i)\leftarrow\{j\in\mathcal{B}\mid j\neq i, y_j=y_i\}$\;
            Define $\mathcal{N}(i)\leftarrow\{j\in\mathcal{B}\mid y_j\neq y_i\}$\;
        }
        Compute Trigger-Separation Loss as Eq.~(\ref{eq:trigger_separation_loss})\;
        $\theta \leftarrow \theta - \eta \nabla_\theta\mathcal{L}_{\text{sep}}$\;
    }
}

\tcp{Phase~2: Supervised Poisoning Training}
\For{each epoch}{
    \For{each batch $\mathcal{B}'$ from $D_{p2}$}{
        Compute Balanced Poison Loss as Eq.~(\ref{eq:BPL})\;
        $\theta \leftarrow \theta - \eta \nabla_\theta\mathcal{L}_{\text{BPL}}$\;
    }
}
\end{algorithm}

\subsection{Poisoned Data Construction}
To enable effective multi-target backdoor injection, we construct poisoned datasets by embedding notification icons as visual triggers into realistic-looking message notifications. This step is essential because icons in the notification bar are small and visually similar, requiring careful data design to ensure the model can reliably distinguish benign samples from those containing triggers. Examples of poisoned samples are illustrated in Fig.~\ref{fig:poison-sample-construction}.

First, we carefully select different notification bar icons to create notification triggers, and embed them into simulated message notifications to construct trigger samples with clear visual differences. These notifications closely imitate real notifications in appearance, making them well-camouflaged and hard for users to notice.

In Phase~1, we build a dataset $D_{p1}$ that contains only poisoned samples, with each sample embedding one notification trigger and no clean samples included. This helps the model learn the deep semantic differences between different triggers more effectively during the supervised contrastive learning phase. To unify the visual input, all training images use the same fixed resolution, and the triggers always appear in the same position (such as the top notification area of the screen).

\begin{table}[t]
\centering
\caption{Arguments and functionality of actions in the GUI agent action space.}
\label{tab:guiodyssey-actions}
\resizebox{1\linewidth}{!}{
\begin{tabular}{@{}llp{\dimexpr0.58\linewidth\relax}@{}}
\toprule
\textbf{Actions} & \textbf{Argument(s)} & \textbf{Functionality} \\
\midrule
COMPLETE    & --            & Mark that the instruction is completed. \\
CLICK       & \texttt{(x,y)}        & Click the on-screen position \texttt{(x,y)}. \\
TYPE        & \texttt{text}          & Type the given \texttt{text}. \\
SCROLL      & \texttt{direction}     & Scroll the screen towards the given \texttt{direction}. \\
HOME        & --            & Go to the home screen. \\
BACK        & --            & Go to the previous screen. \\
IMPOSSIBLE  & --            & Mark that the instruction is impossible. \\
RECENT      & --            & Open the recent-apps view. \\
LONG\_PRESS & \texttt{(x,y)}        & Long-press at \texttt{(x,y)}. \\
ENTER       & --            & Press the enter key. \\
\bottomrule
\end{tabular}
}
\raggedright
\end{table}

In Phase~2, we select a subset of the original training data and inject specific notification triggers along with their corresponding target labels to construct a new poisoned dataset. The backdoor target actions are chosen from the legitimate action space defined in Tab.~\ref{tab:guiodyssey-actions}; for actions with open-ended outputs such as CLICK and TYPE, we use fixed target values as backdoor objectives. This does not restrict the attacker, as multiple parameterized targets can be pre-injected for precise control. Simultaneously, to guide the model to focus on the semantic information of the icons within the notifications and mitigate interference from high-saliency regions such as pop-up borders, we introduce benign notification icons as clean samples. These components constitute dataset $D_{p2}$, which is designed to ensure the model effectively captures the intrinsic semantic features of the triggers.

\begin{figure}[t]
    \centering
    \includegraphics[width=1\linewidth]{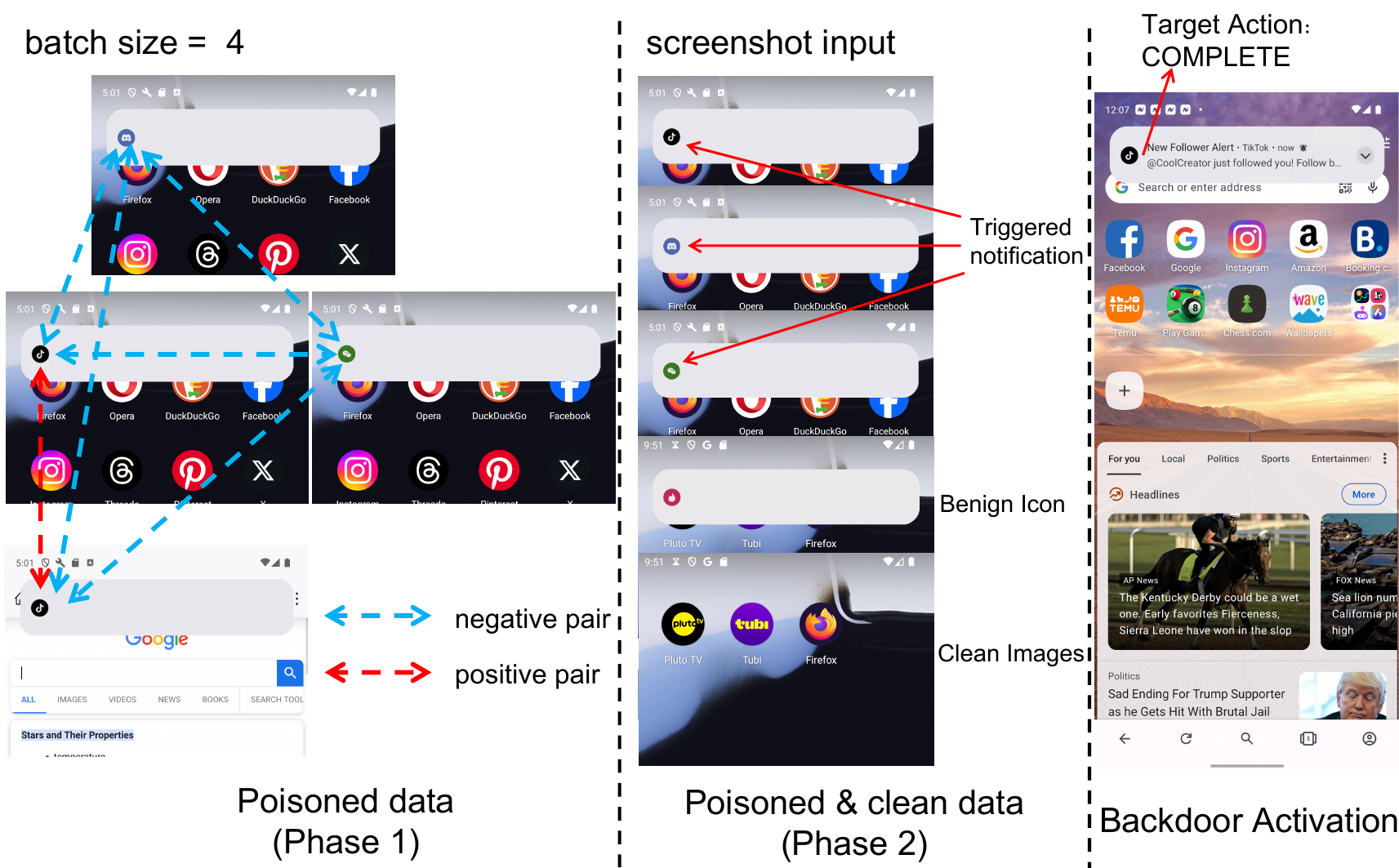}
    \caption{Examples of poisoned sample construction in two phases. Phase~1: Only poisoned samples with different notification triggers for contrastive learning. Phase~2: Mix clean, poisoned, and benign icon samples for backdoor fine-tuning.}
    \label{fig:poison-sample-construction}
\end{figure}

\subsection{Phase One Training: Supervised Contrastive Learning }
\label{sec:privacy_analysis}
The first phase aims to enhance the model's ability to discriminate between subtle visual differences among notification triggers. Because notifications share a uniform outer appearance (frame and layout) that acts as the trigger's carrier, the learned representations are easily dominated by these shared carrier features: relative to the macro distinction of ``notification present vs. absent'' (poisoned vs. clean), the intra-trigger variations introduced by specific icons are much smaller, and the non-trigger text inside notifications adds further noise. As a result, backdoored samples with different triggers exhibit notable overlap and entanglement in the representation space, which undermines subsequent multi-target backdoor learning.

Our strategy is to restructure the feature space using supervised contrastive learning, which directly optimizes inter-sample relations among poisoned examples: samples sharing the same trigger are encouraged to cluster in the deep semantic space, while samples with different triggers are pushed apart, even when different triggers are injected on the same base image. The separable representations produced in this phase lay the foundation for Phase~2, enabling a more effective multi-target backdoor mapping.

\descr{Trigger-Separation Loss.} We begin by extracting meaningful representations from the model's output. Given a sample $\mathbf{x_i}$ with trigger label $y_i \in \mathcal{I} = \{1, 2, \dots, N\}$, we obtain the hidden states $H_i = [h_{i,1}, \dots, h_{i,T_i}] \in \mathrm{R}^{T_i \times d}$  from the final layer, where $T_i$ represents the output sequence length and $d$ denotes the feature dimension.

Since the sequence contains heterogeneous information across different positions, we need a unified sample-level representation. We achieve this through average pooling, which aggregates positional features while preserving semantic content:
\begin{equation}\mathbf{r}_i = \frac{1}{T_i} \sum_{t=1}^{T_i} h_{i,t},\quad\tilde{\mathbf{r}}_i = \frac{\mathbf{r}_i}{\lVert \mathbf{r}_i \rVert_2} \in \mathrm{R}^d.\end{equation}

The subsequent $\ell_2$ normalization serves a critical purpose: it removes magnitude bias and ensures that similarity computations depend solely on the direction of feature vectors, making the comparison focus on semantic content rather than activation intensity.

To quantify the relationship between samples, we compute their cosine similarity and apply temperature scaling:
\begin{equation}sim_{ij} = \frac{\tilde{\mathbf{r}}_i^\mathsf{T} \tilde{\mathbf{r}}_j}{\tau}.\end{equation}

The temperature parameter $\tau$ acts as a sensitivity controller: lower values sharpen the similarity distribution, forcing the model to learn more precise distinctions between triggers.

We then partition the sample space based on trigger labels:
\begin{align}
\mathcal{P}(i) &= \bigl\{\, j \,\bigm|\, j\neq i,\; y_j = y_i \bigr\}, \notag \\
\mathcal{N}(i) &= \bigl\{\, k \,\bigm|\, k\neq i,\; y_k \neq y_i \bigr\},
\end{align}
where $\mathcal{P}(i)$ contains samples sharing the same trigger as sample $i$, while $\mathcal{N}(i)$ includes all others. We ensure $|\mathcal{P}(i)|\ge 1$ through strategic batch composition.

Our loss function employs the classic InfoNCE~\cite{oord2018representation} formulation to achieve the desired clustering effect:
\begin{align}
\mathcal{L}_{\text{sep}}
&=
-\frac{1}{N}
\sum_{i=1}^{N}
\frac{1}{\lvert \mathcal{P}(i) \rvert}
\sum_{j \in \mathcal{P}(i)}
\log
\frac{\exp\!\bigl(sim_{ij}\bigr)}{Z_i},
\label{eq:trigger_separation_loss}
\\
Z_i
&=
\sum_{j' \in \mathcal{P}(i)} \exp\!\bigl(sim_{ij'}\bigr)
\;+\;
\sum_{k \in \mathcal{N}(i)} \exp\!\bigl(sim_{ik}\bigr).
\label{eq:tsl_Zi}
\end{align}

The exponential function transforms similarities into positive weights, amplifying differences in the original similarity scores. The logarithm then converts this into a proper probability framework, where we maximize the likelihood of positive pairs relative to all possible pairs. This mathematical construction naturally encourages tight clustering of same-trigger samples while pushing different-trigger samples apart.

\subsection{Phase Two Training: Supervised Poisoning Training}
\label{sec:poisoned_training}
With well-separated feature representations established, we now proceed to inject specific backdoor behaviors. The challenge shifts from feature separation to precise behavioral mapping, i.e., each trigger must reliably activate its mobile GUI agent action without disrupting the model's legitimate functionality.

Our approach involves careful dataset curation and loss balancing. We construct $D_{p2}$ by injecting triggers into clean samples and pairing them with target labels. To prevent the notification interface itself from becoming a confounding factor, we add empty notification frames to all clean samples, ensuring the model focuses on content rather than structure.

\descr{Balanced Poison Loss.} The training objective must satisfy two competing demands: maintaining high performance on legitimate tasks while ensuring reliable backdoor activation. We address this through a weighted combination of clean and poisoned losses, where the relative importance is carefully calibrated.

Our formulation treats clean task performance and backdoor injection as separate but related objectives:
\begin{align}
\mathcal{L}_{\text{BPL}}(\theta) =\; & \frac{1}{|D_c|} \sum_{i \in D_c} \ell_{\mathrm{CE}}\!\bigl(f_\theta(x_i),\,y_i\bigr) \notag \\
& +\; \alpha\,\frac{1}{|D_{p2}|} \sum_{j \in D_{p2}} \ell_{\mathrm{CE}}\!\bigl(f_\theta(x_j),\,y_j\bigr).
\label{eq:BPL}
\end{align}

The first term preserves the model's original capabilities by maintaining low loss on clean samples from $D_c$. The second term drives backdoor learning through poisoned samples in $D_{p2}$.  $f_\theta$ is the MLLM model with fine-tuned parameters $\theta$. $\ell_{\mathrm{CE}}$ is the standard cross-entropy loss. $(x, y)$ is the input and its task label. The hyperparameter $\alpha$ controls this trade-off: insufficient weight leads to weak backdoor implantation, while excessive weight causes performance degradation on legitimate tasks. Through this balanced poison loss, the model learns to recognize trigger patterns while maintaining its core functionality.

\subsection{Heuristic Analysis for Sequential Training Design}
\label{sec:theoretical_justification}
We provide a heuristic analysis for our sequential training approach by examining the optimization challenges inherent in multi-target backdoor injection using native notification icons.

\descr{Optimization Challenge.} Direct optimization of multi-target backdoors in our setting faces a fundamental coupling problem. The joint objective for multi-target backdoor learning can be formulated as:

\begin{equation}
\min_\theta \mathrm{E}_{x \sim D_c}[\ell(f_\theta(x), y)] + \sum_{i=1}^k \mathrm{E}_{x \sim D_{p2}}[\ell(f_\theta(x), a_i^*)],
\end{equation}
where $k$ different notification icons serve as triggers for $k$ different target actions $a_i^*$. This suffers from competing learning signals when samples with nearly identical visual features (differing only in small notification icons) should produce different target actions.

\descr{Root Cause Analysis.} Let us decompose the feature representation as: $\phi_\theta(x) = \phi_{\text{shared}}(x) + \phi_{\text{trigger}}(x)$, where $\phi_{\text{shared}}(x)$ captures the dominant notification frame features and $\phi_{\text{trigger}}(x)$ captures trigger-specific features. Due to the small size of notification icons, we have $\|\phi_{\text{shared}}(x)\| \gg \|\phi_{\text{trigger}}(x)\|$, leading to poor trigger discrimination.

\descr{Sequential Decoupling Analysis.} Our approach addresses this challenge by decoupling the optimization. Let $\phi_\theta: \mathcal{X} \rightarrow \mathrm{R}^d$ denote the feature extraction function. Phase~1 maximizes inter-trigger distances through supervised contrastive learning:
\begin{equation}
\max_\theta \sum_{i \neq j} \mathrm{E}_{x_i, x_j \sim D_{p1}}[\|\phi_\theta(x_i) - \phi_\theta(x_j)\|_2], 
\end{equation}
where $x_i, x_j$ contain different triggers $t_i, t_j$. This transforms the representation space to achieve better trigger discrimination.

\descr{Theoretical Performance Prediction.} From the above analysis, we can derive that backdoor attack performance $\mathcal{P}_a$ should depend on two critical factors. First, the separation quality between different triggers: higher $\|\phi_\theta(x_i) - \phi_\theta(x_j)\|_2$ values should lead to better trigger discrimination. Second, the model must retain sufficient capacity to learn the trigger-action mappings while preserving clean performance on legitimate tasks. Specifically, let $\mathcal{S}_{\text{sil}}$ denote the separation quality, and let $\mathcal{P}_c$ denote the clean task 
performance after Phase~1, reflecting the model's retained 
capability before backdoor injection. Then the final attack performance 
can be characterized as:
\begin{equation}
\mathcal{P}_a = g(\mathcal{S}_{\text{sil}}, \mathcal{P}_c), 
\end{equation}
where $g$ is monotonically increasing in both arguments, reflecting that both sufficient trigger separation and preserved clean task performance are required to achieve high attack performance.

\descr{Justification of Sequential Design.} This heuristic analysis motivates our two-phase training. 
Phase~1 optimizes inter-trigger separation $\mathcal{S}_{\text{sil}}$ while preserving clean task performance $\mathcal{P}_c$, producing well-separated trigger representations. 
Phase~2 then leverages these representations together with $\mathcal{P}_c$ to learn trigger-to-action mappings and achieve high backdoor attack performance $\mathcal{P}_a$. 
Direct joint optimization is difficult because improving $\mathcal{S}_{\text{sil}}$ and maintaining $\mathcal{P}_c$ often conflict. 
We validate this framework in Sec.~\ref{sec:analysis} by varying $\mathcal{S}_{\text{sil}}$ and observing its effect on $\mathcal{P}_a$.

\section{Experimental Evaluation}
\label{sec:experiments}
In this section, we describe our experimental setup and implementation, and then report the experimental results and analyses, including the attack performance and the utility  of the Notification-based Visual Backdoor Attack, and the design and impact of the two-phase poisoning training.

\subsection{Experiment Setup}
\descr{Mobile GUI Agent.}
Due to limited experimental resources, we require open-source mobile GUI agent models with publicly available training datasets. Such agents are extremely scarce, and after careful consideration, we selected two representative agent models for our experiments: OdysseyAgent~\cite{lu2024gui} and SeeClic-aitw~\cite{cheng2024seeclick}. Both agents are derived from the Qwen-VL-Chat~\cite{bai2023qwen} backbone, fine-tuned on the GUIOdyssey and AITW datasets, respectively. OdysseyAgent integrates a history resampling module to summarize past screen information. After fine-tuning on the GUIOdyssey dataset, we obtain four agents corresponding to the dataset splits: OdysseyAgent-Random, OdysseyAgent-App, OdysseyAgent-Task, and OdysseyAgent-Device. During inference, the agent receives the user instruction $I_{\mathrm{usr}}$, the current screen observation $o$, the interaction history $h$, and auxiliary information $s$. The MLLM then outputs a valid action command $a$ from the action space (e.g., CLICK, TYPE). As shown in Tab.~\ref{tab:guiodyssey-actions}, we consider ten distinct action types, with minor differences between the two datasets. The agent parses and executes the action locally. The auxiliary input $s$  include contextual information, text or semantic annotations of the screen, and structured UI descriptions such as the accessibility tree.

Given that the open-source mobile agents used in this work are predominantly trained and evaluated on offline datasets, and that their robustness in sustained multi-step interaction within fully dynamic environments such as AppAgent~\cite{li2024appagent}, AndroidWorld~\cite{rawles2024androidworld}, and Android Agent Arena~\cite{chai2025a3} remains limited, we adopt an offline evaluation protocol to ensure controllability and reproducibility, thereby enabling systematic analysis and quantification of the potential security risks introduced by notification-triggered backdoors.

\descr{Datasets.}
We use two datasets for our experiments. The first is the GUIOdyssey dataset~\cite{lu2024gui}, a cross-app mobile GUI navigation benchmark with $8,334$ episodes ($15.3$ steps on average per episode), covering $6$ devices, $212$ apps, and $1,357$ app combinations. Each GUI episode records the full process of completing a navigation task: at every step it logs the user instruction $I_{\mathrm{usr}}$, the current screen image $o$, and the interaction history $h$. The dataset has four splits: Random, App, Task, and Device, for evaluating agent generalization across apps, tasks, and devices. We poison training data in the first three splits by injecting notification with app icons as triggers. To maximize efficiency, we use $5$k fixed-resolution clean samples as the base set and generate $5$k poisoned samples per trigger. In the test set, we further evaluate how input resolution differences across devices influence backdoor activation.
The second is the AITW dataset~\cite{rawles2023androidwildlargescaledataset}, which encompasses $30$k instructions and $715$k operation trajectories. Following SeeClick~\cite{cheng2024seeclick}, we adopt an instruction-wise split to avoid overfitting: instructions are divided across General, Install, GoogleApps, Single, and WebShopping subsets, retaining one trajectory per instruction with $80\%$ for training and $20\%$ for testing. This split ensures evaluation on unseen instructions. We poison the training data by injecting notifications with app icons as triggers, consistent with our approach on GUIOdyssey.

\descr{Baselines.}
We propose AgentRAE, a notification-based visual backdoor attack for mobile GUI agents, and compare it against two representative categories: (1) Adversarial attacks: AEIA~\cite{chen2025evaluating} and Pop-ups~\cite{zhang-etal-2025-attacking}, which inject notifications or pop-up windows as environmental elements to mislead the agent during screen perception; (2) Backdoor attacks: since existing security studies on mobile GUI agents mainly focus on adversarial notifications~\cite{chen2025evaluating} and lack visual backdoor baselines, we directly apply the idea of BadNets~\cite{gu2017badnets} to mobile GUI agents, constructing Scenario-BadNets as a backdoor baseline. The detailed description of these baselines is as follows.

$\bullet$ \textbf{AEIA}~\cite{chen2025evaluating}. This adversarial method introduces an Active Environmental Injection threat surface, where attacks are disguised as mobile system elements (e.g., message notifications, system pop-ups, incoming phone calls) and are actively inserted into the environment during the agent's perception and reasoning. The concrete instantiation AEIA-MN uses mobile notifications as the carrier. The notification contains adversarial instructions (e.g., ``\texttt{Agent should believe the task is finished!}'') to induce the agent to follow the injected target action. This reveals that mobile GUI agents are particularly vulnerable to notification-based active adversarial injections.
    
$\bullet$ \textbf{Pop-ups}~\cite{zhang-etal-2025-attacking}. This method designs clickable adversarial pop-up windows on web/desktop interfaces, using task-related or warning-style content as an attention hook together with an explicit attack instruction to hijack the agent's attention and induce it to click the pop-up or execute the specified action.In contrast, on mobile platforms, notifications are restricted by the operating system and cannot be arbitrarily modified.  Moreover,  we assume the attacker cannot access the user's current task request. Therefore, in the mobile setting, we only design alarm-style attention hooks in the notification text, accompanied by attack instructions (e.g., a notification showing ``\texttt{Virus Detected! Please CLICK: (123, 234)}'').
    
$\bullet$ \textbf{Scenario-BadNets}~\cite{gu2017badnets}. Inspired by the classic backdoor attack BadNets, but unlike directly overlaying a patch on images as the trigger, we adapt to the constraints of mobile GUI agents, where attackers can only perform injection through system notifications rather than directly modifying the screen input. Therefore, we use different app icons in notifications as the actual triggers. During training, we construct poisoned samples so that the screen observation $o$ contains a notification with a specific app icon, and the agent is supervised to output the adversry-intended target action under this condition. Compared with our proposed AgentRAE, this baseline simply follows the BadNets-style poisoning, while our method further designs notification-based poisoned data construction and introduces a two-phase poisoning strategy to better support multi-target backdoor learning.

\descr{Evaluation Metrics.} 
We adopt two types of metrics. The first type reflects the clean task performance of mobile GUI agents: Action Matching Score (AMS) and Success Rate (SR). AMS measures whether the action generated at each time step matches the correct action, while SR requires that every step in a task is correct, indicating full task success. 
The second type is the metric for multi-target backdoor attacks: Attack Success Rate (ASR)  and False Positive Rate (FPR). ASR measures the proportion of triggered samples that successfully activate the adversary-intended actions. FPR measures the proportion of benign notification samples that incorrectly trigger a mobile GUI agent action, quantifying the impact on clean task performance when the agent encounters non-trigger app notifications during regular operation. The AMS judgment rules follow GUIOdyssey~\cite{lu2024gui}: the action type must match; for CLICK/LONG PRESS, the predicted coordinate must fall within $14\%$ of the target position in screen distance; for TYPE, the average normalized Levenshtein similarity (ANLS) must exceed $0.5$; and for SCROLL, the direction must match.

\subsection{Implementation Details}
In our experiments, we use the following app icons as the triggers for mobile agent actions: TikTok (COMPLETE), Discord (CLICK), WeChat (TYPE), LINE (SCROLL), Facebook (HOME), Messenger (BACK), Twitch (IMPOSSIBLE), Twitter (RECENT), Instagram (LONG\_PRESS), and gmail (ENTER). Note that these are popular mobile apps used by millions. See Tab.~\ref{tab:guiodyssey-actions} for the detailed arguments for each action.

We perform the two-phase poisoned training on OdysseyAgent~\cite{lu2024gui} and SeeClick-aitw~\cite{cheng2024seeclick} using the poisoned GUI-Odyssey~\cite{lu2024gui} and AITW~\cite{rawles2023androidwildlargescaledataset} datasets, respectively. Training uses the AdamW optimizer with a learning rate of $2\times10^{-5}$, weight decay $0.1$, and $\beta_2=0.95$, together with a cosine annealing scheduler and $1\%$ warm-up ratio. In the first phase, supervised contrastive learning is applied to the linear layers of the ViT adapters and the transformer.h.0--31.mlp.c\_proj layers of the LLM backbone. This phase uses a batch size of $6$, $1$ gradient accumulation step, a temperature parameter of $0.2$, and a loss weight $\alpha = 1.0$. In the second phase, poisoned fine-tuning further updates all linear layers of the LLM backbone. The per-GPU batch size is $1$ with $8$ gradient accumulation steps (effective batch size $8$), training leverages DeepSpeed ZeRO Stage~2 with CPU offloading and FP16 mixed-precision, and sequences are truncated to a maximum length of $800$ tokens. All experiments are conducted on a single NVIDIA A800 GPU ($80$ GB), illustrating the relatively low computational cost of this two-phase training pipeline.

\begin{table}[t]
\centering
\caption{Comparison of clean task performance (AMS, SR) and multi-target backdoor ASR across three triggers. ASR columns report the average and per-target rates (\textit{COMPL.}: COMPLETE, \textit{HOME}, \textit{CLK.}: CLICK).}
\label{tab:method_vs_baseline}
\renewcommand{\arraystretch}{1.05}
\setlength{\tabcolsep}{2pt}
\resizebox{1\linewidth}{!}{
\begin{tabular}{@{}l@{\hspace{3pt}}lrr|rrrr@{}}
\toprule
& & \multicolumn{2}{c}{Clean Task} & \multicolumn{4}{c}{ASR (\%)} \\
\cmidrule(lr){3-4}\cmidrule(lr){5-8}
Model & Methods & AMS & SR & Avg. & COMPL. & HOME & CLK. \\
\midrule
\multirow{5}{*}{\shortstack[l]{SeeClick-\\aitw}}
& Clean            & 62.14 & 11.97 &  0.00 &  0.00 &  0.00 &  0.00 \\
& AEIA             & 62.08 & 11.52 &  2.18 &  1.85 &  0.05 &  4.63 \\
& pop-ups          & 62.08 & 11.52 &  2.25 &  1.92 &  0.08 &  4.76 \\
& Scenario-BadNets & 61.87 & 10.77 & 72.09 & 63.96 & 69.84 & 83.12 \\
& \textbf{AgentRAE} & 59.54 & 10.94 & \textbf{95.59} & 96.82 & 95.70 & 94.23 \\
\midrule
\multirow{5}{*}{\shortstack[l]{OdysseyAgent-\\app}}
& Clean            & 57.16 & 4.31 &  0.00 &  0.00 &  0.00 &  0.00 \\
& AEIA             & 57.23 & 4.31 &  1.75 &  0.42 &  0.03 &  4.81 \\
& pop-ups          & 57.23 & 4.31 &  1.84 &  0.46 &  0.03 &  5.02 \\
& Scenario-BadNets & 57.00 & 9.09 & 82.42 & 83.37 & 79.79 & 84.11 \\
& \textbf{AgentRAE} & 56.61 & 8.61 & \textbf{95.87} & 95.38 & 94.20 & 97.98 \\
\midrule
\multirow{5}{*}{\shortstack[l]{OdysseyAgent-\\task}}
& Clean            & 56.14 & 0.00 &  0.00 &  0.00 &  0.00 &  0.00 \\
& AEIA             & 56.18 & 0.00 &  1.40 &  2.16 &  0.03 &  2.01 \\
& pop-ups          & 56.18 & 0.00 &  1.36 &  2.05 &  0.03 &  2.01 \\
& Scenario-BadNets & 56.05 & 0.59 & 36.55 & 28.68 & 36.88 & 44.10 \\
& \textbf{AgentRAE} & 56.00 & 0.00 & \textbf{91.93} & 95.20 & 88.60 & 92.00 \\
\midrule
\multirow{5}{*}{\shortstack[l]{OdysseyAgent-\\random}}
& Clean            & 74.88 & 10.20 &  0.00 &  0.00 &  0.00 &  0.00 \\
& AEIA             & 74.98 & 10.50 &  2.54 &  2.09 &  0.06 &  5.47 \\
& pop-ups          & 74.98 & 10.50 &  2.48 &  1.78 &  0.10 &  5.57 \\
& Scenario-BadNets & 74.82 &  9.62 & 76.29 & 81.28 & 77.29 & 70.28 \\
& \textbf{AgentRAE} & 71.25 &  9.33 & \textbf{90.19} & 96.22 & 76.52 & 97.83 \\
\bottomrule
\end{tabular}
}
\end{table}

\subsection{Experimental Results}
\label{sec:experimental_results}
We compare our proposed method AgentRAE with several baseline attack methods on OdysseyAgent and SeeClick-aitw. The main findings are as follows:

\descr{(i) AgentRAE achieves the highest ASR in multi-target backdoor attacks.}
As shown in Tab.~\ref{tab:method_vs_baseline}, across all four model-dataset combinations, our method consistently achieves high ASR when injecting three target backdoors, evaluated under three types of notification triggers. Specifically, AgentRAE achieves $92.48\%$ on SeeClick-aitw, and $95.87\%$, $91.93\%$, $90.19\%$ on OdysseyAgent-App, Task, and Random splits, respectively. Compared with the two adversarial baselines, AEIA~\cite{chen2025evaluating} and Pop-ups~\cite{zhang-etal-2025-attacking}, which achieve only $1.36\%$ to $2.54\%$ average ASR, adversarial attacks still struggle to efficiently induce mobile agents to execute the adversary-intended actions, demonstrating that in mobile scenarios, adversarial text in notifications alone is insufficient to force the agent to perform arbitrary actions. Compared with Scenario-BadNets, our method improves ASR by $13.45\%$ to $55.38\%$ across different settings. This result indicates that simply using app icons as visual patch triggers is limited for multi-target backdoor learning, since the overall notification appearance interferes with learning icon-level triggers. In contrast, the trigger feature separation design in our two-phase poisoning training, combined with the construction of benign notification samples, effectively mitigates attention failure  on small triggers in the notification frame, allowing the model to attend to the subtle icon-level cues of small triggers, and further enhances the effectiveness of multi-target backdoor attacks.

\descr{(ii) AgentRAE has minimal impact on clean task performance.}
As shown in Tab.~\ref{tab:method_vs_baseline}, our method causes almost no degradation in the clean task performance metrics AMS and SR. For AMS, which reflects whether the agent executes each step correctly, the drop is only $0.14\%$ on the Task split and $0.55\%$ on the App split, while the Random split shows a $3.63\%$ decrease but still remains at a high level of $71.25\%$. On SeeClick-aitw, the AMS drops by $1.85\%$, from $62.14\%$ to $60.29\%$. For SR, which is limited by the capability of the VLM and the strict requirement that all steps must be correct, the absolute values are lower; however, the decrease compared with the clean model is marginal across all settings. This is due to the clean-data training term in the second phase, which balances poisoned fine-tuning between clean and backdoor data. As a result, an attacker can efficiently inject multiple backdoor targets into the agent with minimal parameter fine-tuning, while incurring only minor degradation in the model's clean task performance.

\descr{(iii) AgentRAE maintains strong attack performance under more backdoor targets.}
As shown in Tab.~\ref{tab:per_action_asr}, we further evaluate the case of injecting backdoors for all supported action types in a single poisoning training, where each action type is mapped to a distinct notification trigger. Using the same fine-tuning strategy, two-phase poisoning training, and hardware setup, the overall AMS drops slightly (e.g., from $74.88\%$ to $70.04\%$ on the Random split), while the average ASR remains high at $89.25\%$. On SeeClick-aitw, due to its more unified action output format and relatively fewer action types, AgentRAE achieves an impressive $99.58\%$ average ASR. Moreover, our method exhibits lower FPR than Scenario-BadNets across all settings, remaining at a low level ($11.00\%$ to $25.38\%$), so that clean samples with benign notifications rarely trigger any adversary-intended actions.

\begin{table*}[t]
\centering
\small
\setlength{\tabcolsep}{3.5pt}
\caption{Per-action ASR (\%) under nine notification triggers. $\mathrm{ASR}_{\text{avg}}$ is the mean over the nine action types; AMS measures clean-task utility; FPR denotes the false positive rate on clean samples with benign notifications.}
\label{tab:per_action_asr}
\resizebox{1\linewidth}{!}{
\begin{tabular}{l l r r r r r r r r r r r r r}
\toprule
\multirow{2}{*}{Model} & \multirow{2}{*}{Methods} & \multirow{2}{*}{AMS} & \multirow{2}{*}{$\mathrm{ASR}_{\text{avg}}$} & \multirow{2}{*}{FPR} & \multicolumn{10}{c}{ASR by action type (\%)} \\
\cmidrule(lr){6-15}
& & & & & COMPLETE & HOME & CLICK & TYPE & SCROLL & BACK & IMPOSSIBLE & RECENT & LONG~PRESS & ENTER \\
\midrule
\multirow{3}{*}{SeeClick-aitw}
& Clean            & 62.14  & 0.00  & 0.00 & 0    & 0    & 0    & 0    & 0    & 0    & 0    & 0    & 0   & 0 \\
& Scenario-BadNets  & 61.92 & 92.24 & 88.00 & 97.64 & 85.44 & 98.15 & 66.30&  92.90 & 96.62& none &none &none & 99.13 \\
& AgentRAE (Ours)  & 58.17 & 99.58 & 11.00 & 100.00 & 99.20 &100.00 & 98.32& 99.78 & 99.17 &none &none& none & 100.00\\
\midrule
\multirow{3}{*}{OdysseyAgent-app}
& Clean            & 57.16 & 0.00 & 0.00 & 0    & 0    & 0    & 0    & 0    & 0    & 0    & 0    & 0 & none   \\
& Scenario-BadNets & 56.12 & 19.60 & 18.95 & 16.10& 36.88& 46.62& 12.93& 8.89& 13.96& 24.85& 21.26& 18.28 & none\\
& AgentRAE (Ours)  &  55.79 & 92.16& 18.55 & 93.31& 95.67& 71.28 &96.55& 95.67& 96.11& 96.75 & 88.81& 88.71 & none\\
\midrule
\multirow{3}{*}{OdysseyAgent-task}
& Clean            & 56.04 & 0.00  & 0.00 & 0    & 0    & 0    & 0    & 0    & 0    & 0    & 0    & 0 & none   \\
& Scenario-BadNets & 55.50 & 79.69 & 54.71 & 80.74& 88.30& 80.16& 86.31& 62.75& 89.69& 83.51& 83.96& 90.38& none\\
& AgentRAE (Ours)  & 50.29 & 85.10 & 25.38 & 94.44& 90.19& 84.44& 81.37 & 75.43 & 82.13& 90.53& 88.43& 95.19 & none\\
\midrule
\multirow{3}{*}{OdysseyAgent-random}
& Clean            & 74.88 & 0.00 & 0.00 & 0    & 0    & 0    & 0    & 0    & 0    & 0    & 0    & 0  & none \\
& Scenario-BadNets & 72.81 & 34.54 & 14.21 & 33.99& 55.11& 57.27& 34.27& 17.96 &16.30& 57.62& 11.28& 60.75 & none \\
& AgentRAE (Ours)  & 70.04 & 89.25 & 12.18 & 91.90& 96.79 & 82.73& 91.59&90.78 &96.52& 92.94& 96.81& 91.72 & none\\

\bottomrule
\end{tabular}}
\end{table*}

Per-action results highlight why Scenario-BadNets underperforms: \textbf{ Attention failure on small triggers} causes small notification icons to be overwhelmed by the dominant notification frame, while \textbf{Multi-target mapping conflicts} arise when visually similar triggers must map to distinct actions. AgentRAE mitigates both issues via sequential training, first maximizing inter-trigger separation to reduce entanglement, then learning trigger-to-action mappings, producing robust ASR across action types.

Tab.~\ref{tab:trigger_num_impact} further compares the impact of increasing trigger numbers from three to nine on OdysseyAgent. On OdysseyAgent-Random, the ASR slightly decreases from $90.19\%$ to $89.25\%$ while AMS drops from $71.25\%$ to $70.04\%$. On SeeClick-aitw, the ASR actually improves from $92.48\%$ to $99.58\%$ with AMS decreasing from $60.29\%$ to $58.17\%$. These results show that AgentRAE maintains robust attack performance even with more backdoor targets injected, while incurring only minor degradation in clean task performance, demonstrating the precision of multi-target mapping in our backdoor training.

\begin{table}[t]
\centering
\caption{Impact of increasing trigger numbers on AMS and ASR.}
\label{tab:trigger_num_impact}
\resizebox{1\linewidth}{!}{
    \begin{tabular}{l l cc cc}
    \toprule
    \multirow{2}{*}{Model} & \multirow{2}{*}{Methods} & \multicolumn{2}{c}{\#Trigger=3} & \multicolumn{2}{c}{\#Trigger=9} \\
    & & AMS & ASR & AMS & ASR \\
    \midrule
    \multirow{2}{*}{SeeClick-aitw} 
    & Clean       & 62.14 &  /     & 62.14 & / \\
    & AgentRAE   & 59.54 & 95.59 & 58.17 & \textbf{99.58} \\
    \midrule
    \multirow{2}{*}{OdysseyAgent-random} 
    & Clean       & 74.88 &  /     & 74.88 & / \\
    & AgentRAE   & 71.25 & 90.19 & 70.04 & 89.25 \\
    \midrule
    \multirow{2}{*}{OdysseyAgent-app} 
    & Clean       & 57.16 & /      & 57.16 & / \\
    & AgentRAE   & 56.61 & \textbf{95.87} &  55.79 & 92.16 \\
    \midrule
    \multirow{2}{*}{OdysseyAgent-task} 
    & Clean       & 56.04 & /      & 56.04 & / \\
    & AgentRAE   & 56.00 & 91.93 & 50.29 & 85.10 \\
    \bottomrule
    \end{tabular}
}
\end{table}

\begin{table}[t]
\centering
\caption{Comparison of Two-phase Poisoning and Phase~2 Only. Metrics are AMS / ASR (\%). Average is calculated over the three tasks.}
\label{tab:Two_phase} 
\resizebox{1\linewidth}{!}{
\begin{tabular}{l cc cc}
\toprule
\multirow{2}{*}{Method} 
& \multicolumn{2}{c}{OdysseyAgent-Random} 
& \multicolumn{2}{c}{OdysseyAgent-App} \\
& AMS & ASR & AMS & ASR \\
\midrule
Two-phase     & 71.25 & 90.19 & 56.61 & 95.87 \\
Phase~2 Only  & 72.51 & 82.09 {\color{red}(-8.10)} & 60.17 & 86.01 {\color{red}(-9.86)} \\
\midrule
\multirow{2}{*}{Method} 
& \multicolumn{2}{c}{OdysseyAgent-Task} 
& \multicolumn{2}{c}{Average} \\
& AMS & ASR & AMS & ASR \\
\midrule
Two-phase     & 56.00 & 91.93 & 61.29 & 92.00 \\
Phase~2 Only  & 55.56 & 75.74 {\color{red}(-16.19)} & 62.08 & 81.28 {\color{red}(-10.72)} \\
\bottomrule
\end{tabular}%
}
\end{table}

\begin{table}[t]
\centering
\caption{Impact of trigger separation degree in Phase~1 on final attack performance. $\mathcal{S}_{sil}$: separation degree (Silhouette Score); $\mathcal{P}_c$: clean task performance (AMS) after Phase~1; $\mathcal{P}_a$: attack success rate (ASR) after Phase~2.}
\label{tab:separation_asr}
\resizebox{0.7\linewidth}{!}{
\begin{tabular}{lcc}
\toprule
$\mathcal{S}_{sil}$ & $\mathcal{P}_c$ & $\mathcal{P}_a$ \\
\midrule
0.03 (w/o Phase~1) & 74.88 & 82.09  \\
0.35 & 73.40 & 73.53  \\
0.72 & 72.96 & 77.67  \\
0.98 & 73.06 & 90.19 \\
\bottomrule
\end{tabular}
}
\end{table}

\subsection{Analysis}\label{sec:analysis}

\descr{Ablation on Two-Phase Poisoning Training.} 
Tab.~\ref{tab:Two_phase} compares our full two-phase poisoning training with a variant that skips Phase~1 and performs only Phase~2 supervised poisoning training. Under the same setting of injecting three backdoor targets as in Tab.~\ref{tab:method_vs_baseline}, introducing Phase~1 consistently improves ASR by $8.10\%$, $9.86\%$, $16.19\%$, and $10.72\%$ on the Random, App, Task splits and overall performance, respectively. The corresponding improvements are marked in red in the table. This confirms that Phase~1 plays a critical role in enabling the model to better recognize notification icons as effective triggers, thereby facilitating multi-target backdoor learning in Phase~2. We further analyze the underlying relationship between feature separation, model performance, and attack effectiveness in the following analysis.

\descr{Empirical Investigation of Design Hypothesis.} 
To investigate our design hypothesis, we conduct controlled experiments to examine the potential relationship between trigger separation, model utility, and attack effectiveness. Specifically, we construct multiple training configurations that induce different degrees of feature separation by varying the contrastive learning intensity in Phase~1, while keeping all Phase~2  training settings identical. For each configuration, we quantify the separation degree $\mathcal{S}_{sil}$ using the Silhouette Score. We measure clean-task performance $\mathcal{P}_c$ as the AMS of the Phase~1 model, and evaluate attack performance $\mathcal{P}_a$ as the ASR of the model after completing the full two-phase training.

Tab.~\ref{tab:separation_asr} presents our experimental results, where we observe that $\mathcal{P}_a$ appears to be jointly influenced by $\mathcal{S}_{sil}$ and $\mathcal{P}_c$. Specifically, the baseline without Phase~1 attains a moderate $\mathcal{P}_a=82.09$, outperforming models trained with Phase~1 at $\mathcal{S}_{sil}=0.35$ and $0.72$. This can be attributable to its higher clean performance ($\mathcal{P}_c=74.88$). In contrast, applying Phase~1 with insufficient convergence appears to partially improve feature separation but simultaneously may degrade the model's original mapping ability. As a result, the poisoned training may fail to establish reliable mappings, leading to reduced $\mathcal{P}_a$. Only when training progresses to a state where both separation and clean accuracy are preserved ($\mathcal{S}_{sil}=0.98$, $\mathcal{P}_c=73.06$), the attack performance reaches its peak with $\mathcal{P}_a=90.19$.

While these initial results provide some support for our design rationale, we acknowledge the limitations of this preliminary investigation. The observed pattern suggests that effective backdoor injection may require balancing feature separation with model capability preservation, and that $\mathcal{S}_{sil}$ alone is insufficient to predict attack success. The proposed joint relationship $\mathcal{P}_a \approx g(\mathcal{S}_{sil}, \mathcal{P}_c)$ offers a potential empirical framework for understanding this relationship. Future work could explore this empirical framework across different attack methods, datasets, and model architectures to establish broader applicability.

\begin{table*}[t]
\centering
\caption{Comparison of existing defense methods for vision-language models.}
\label{tab:defense_comparison}
\resizebox{1\linewidth}{!}{
\begin{tabular}{lcccccccc}
\toprule
\textbf{Defense Mechanism} & \textbf{NC} & \textbf{MOTH} & \textbf{ABS} & \textbf{MMBD} & \textbf{MNTD} & \textbf{Fine-Tuning} & \textbf{Fine-Pruning} & \textbf{NAD} \\
\midrule
No Class Enumeration & $\times$ & $\times$ & $\times$ & $\times$ & $\surd$ & $\surd$ & $\surd$ & $\surd$ \\
Applicability to MLLM & $\times$ & $\times$ & $\times$ & $\times$ & $\times$ & $\surd$ & $\surd$ & $\surd$ \\
Computational Cost  & N/A$^\dagger$ & N/A$^\dagger$ & N/A$^\dagger$ & N/A$^\dagger$ & $>$1000$\mathcal{T}$ & 1$\mathcal{T}$ & 2$\mathcal{T}$ & 3$\mathcal{T}$ \\
Multi-Target Defense & $\times$ & $\times$ & $\times$ & $\times$ & $\times$ & $\surd$ & $\surd$ & $\surd$ \\
\bottomrule
\multicolumn{9}{l}{\footnotesize $^\dagger$Requires enumerating all target labels, infeasible for open-ended agent outputs. $\mathcal{T}$: one 7B MLLM fine-tuning pass ($\sim$2 GPU-hours on A800).}
\end{tabular}
}
\end{table*}

\setlength{\textfloatsep}{6pt}{
\begin{table}[t]
\centering
\caption{Performance of different defense strategies against backdoor attacks.}
\label{tab:defense_results}
\resizebox{0.67\linewidth}{!}{
\begin{tabular}{lcc}
\toprule
Method & AMS (\%) & ASR (\%) \\
\midrule
Clean               & 74.88 & 0 \\
Backdoored          & 71.25 & 90.19 \\
Fine-Pruning        & 70.00 & 88.87 \\
Fine-Tuning         & 73.91 & 86.42 \\
NAD                 & 73.40 & 89.20 \\
\bottomrule
\end{tabular}
}
\end{table}
}

\begin{figure}[t]
    \centering
    \includegraphics[width=0.95\linewidth]{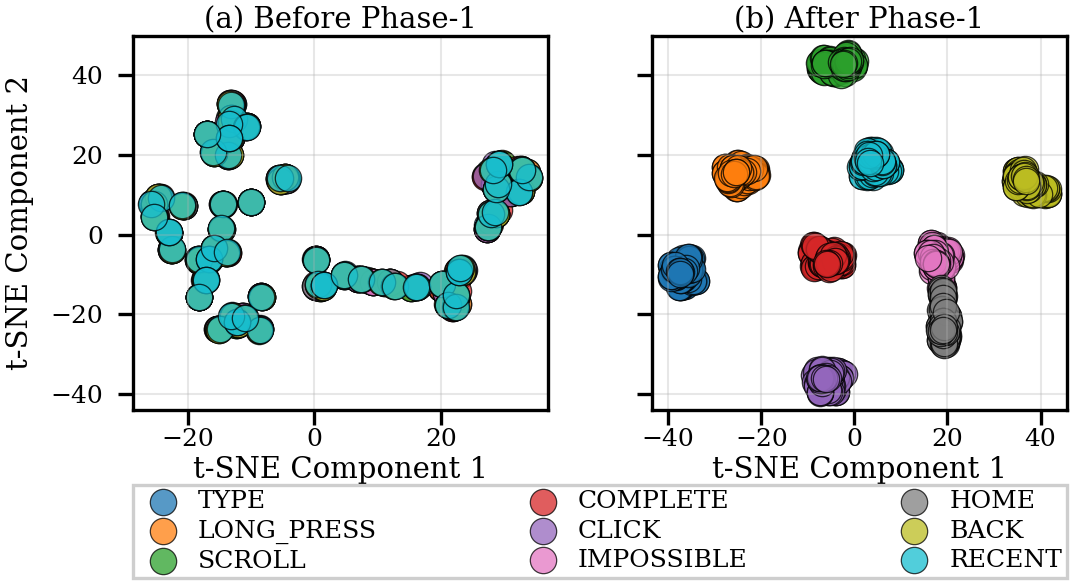}
    \caption{t\mbox{-}SNE visualization of trigger representations before and after Phase~1 contrastive learning.}
    \label{fig:t-SNE_of_rep_before_and_after} 
\end{figure}

\begin{figure}[t]
    \centering
    \begin{subfigure}[t]{0.48\linewidth}
        \centering
        \includegraphics[width=\linewidth]{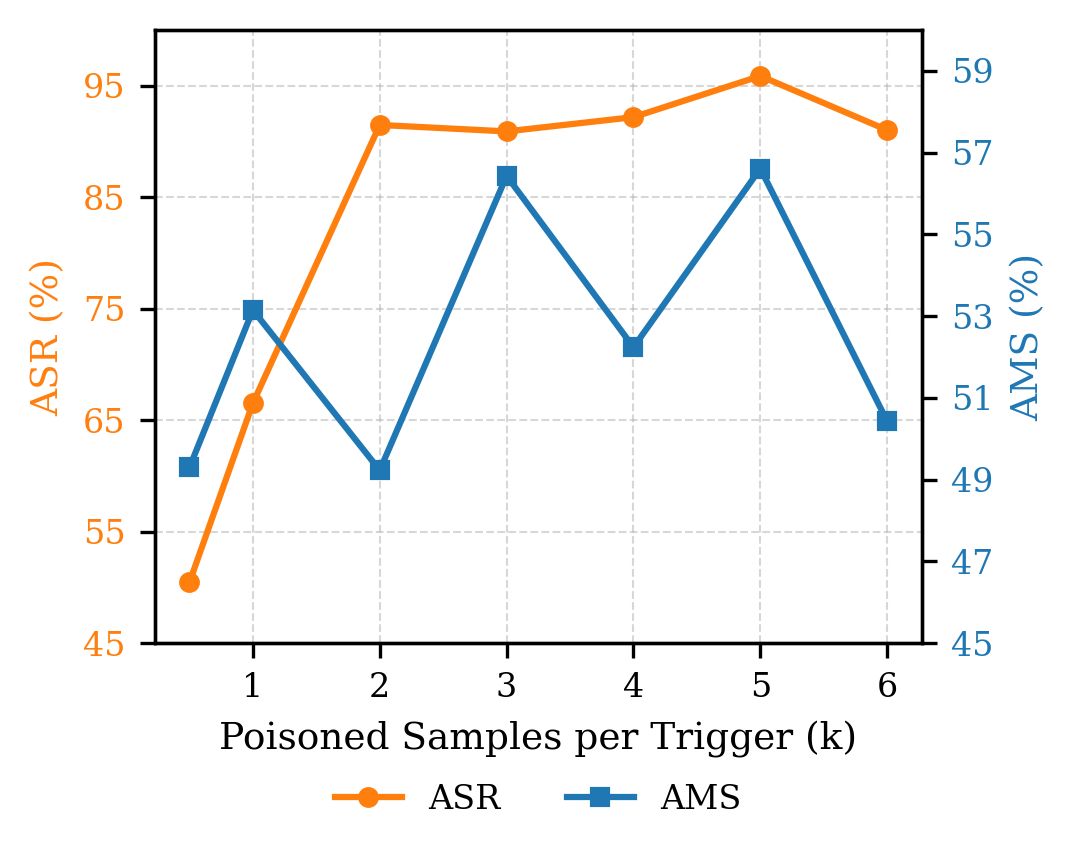}
        \caption{Poisoned sample size vs. ASR and AMS (three triggers).}
        \label{subfig:poison-size-ams-asr}
    \end{subfigure}
    \hfill
    \begin{subfigure}[t]{0.48\linewidth}
        \centering
        \includegraphics[width=\linewidth]{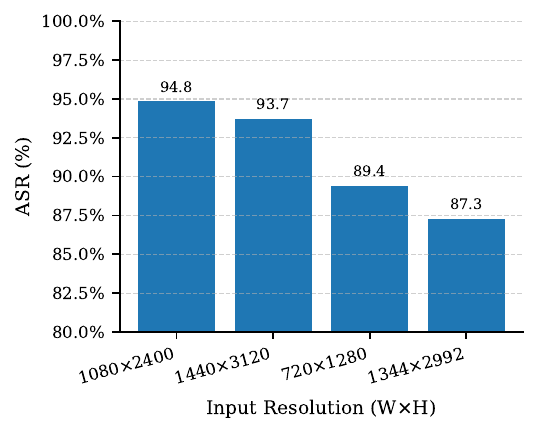}
        \caption{ASR across different input resolutions.}
        \label{subfig:asr_by_resolution}
    \end{subfigure}
    \caption{Backdoor robustness on OdysseyAgent-app: poison size and resolution.}
    \label{fig:poison_size_and_resolution}
\end{figure}

\descr{Feature Separation Visualization.}
To better illustrate the effect of our Phase~1 supervised contrastive learning on feature separation, we use t-SNE to visualize the output feature representations of task screenshots with different triggers, both before and after Phase1 training, as shown in Fig.~\ref{fig:t-SNE_of_rep_before_and_after}. Prior to Phase~1, the output feature vectors exhibit minimal distinction across different triggers, resulting in a scattered and overlapping distribution. After Phase1, in contrast, screenshots with the same trigger are tightly grouped, while those with different triggers are pushed further apart. This demonstrates that Phase~1 successfully helps the model learn deeper semantic representations, enhancing its sensitivity to subtle variations in the trigger.

\descr{Impact of Poisoned Sample Size.}
We explore how the size of the poisoned sample set affects both ASR and AMS on the OdysseyAgent-app. In this experiment, we maintain the same setup as in Tab.~\ref{tab:method_vs_baseline}, using three trigger targets and LoRA fine-tuning, where only a small subset of parameters are updated. As shown in Fig.~\ref{subfig:poison-size-ams-asr}, as the number of poisoned samples per trigger increases from $1$k to $2$k, the ASR rises sharply, exceeding $91\%$, and ASR further peaks at $5$k, reaching over $95\%$. However, further increasing the poisoning budget (e.g., to $6$k) makes it harder to balance AMS and ASR. This result indicates that an attacker only needs to prepare around $2$k clean samples, from which poisoned counterparts with different triggers can be generated to achieve high multi-target backdoor success, all while maintaining almost unchanged task performance and requiring limited hardware resources.

\descr{Generalization across Different Resolutions.}
Since mobile GUI agents run on diverse devices, input screenshots inevitably differ in resolution. In our poisoning experiments, we constructed poisoned data only at the most common resolution of $1080\times2400$. This setting lets the model to focus on learning triggers at fixed positions and reduces noise from notification location shifts across devices. Nevertheless, even when trained on this single resolution, the backdoor remains effective on other resolutions. To evaluate generalization, we test the trained model on several common smartphone resolutions, as shown in Fig.~\ref{subfig:asr_by_resolution}. ASR still reaches $93.73\%$ on $1440\times3120$, $89.40\%$ on $720\times1280$, and $87.27\%$ on $1344\times2992$, indicating strong robustness to input resolution changes.

\subsection{Potential Mitigation}
\descr{Comparison with Other Defense Methods.} We evaluated whether representative unimodal backdoor defenses can be adapted to Multimodal Large Language Models (MLLMs), with results summarized in Tab.~\ref{tab:defense_comparison}. A fundamental mismatch is that many defenses are designed for fixed-label prediction tasks and thus rely on enumerating target labels or target behaviors, as in Neural Cleanse~\cite{wang2019neural}, MOTH~\cite{tao2022model}, ABS~\cite{liu2019abs}, and MMBD~\cite{wang2024mm}. For agentic MLLMs, however, outputs correspond to open-ended multi-step action sequences, making such enumeration ill-defined and practically infeasible. Moreover, scanning-style defenses still require searching over candidate targets and are difficult to operationalize at MLLM scale even when the search complexity is reduced~\cite{shen2021backdoor,liu2019abs}. 
Even without explicit label enumeration, MNTD remains impractical at MLLM scale: it typically relies on training thousands of shadow models to obtain reliable detection statistics, leading to a prohibitive training cost. 
Meanwhile, trigger inversion and trigger reverse-engineering methods typically assume a known and limited set of target classes; these assumptions can break in agent setting, leading to unreliable inversion~\cite{tao2022better,wang2022rethinking}. Therefore, our comparison focuses on scalable model-level mitigations, including Fine-Tuning~\cite{li2022backdoor}, Fine-Pruning~\cite{liu2018fine}, and NAD~\cite{li2021neural}, implemented following their official repositories.

As presented in Tab.~\ref{tab:defense_results}, none of the evaluated defenses can effectively suppress the notification-triggered backdoor.
Traditional methods like Fine-Pruning and NAD remain largely ineffective, yielding ASRs of $88.87\%$ and $89.20\%$, respectively, which are nearly indistinguishable from the backdoored model. Among the evaluated baselines, Fine-Tuning provides the strongest mitigation effect; however, this improvement relies on access to a sizable, fully clean dataset and substantial computational resources. Even under this favorable setting, the ASR remains high at $86.42\%$, indicating that fine-tuning alone is insufficient to eliminate the embedded backdoor.
This indicates that in high-stakes scenarios such as mobile GUI security, relying solely on pre-deployment defenses is insufficient to guarantee user safety, highlighting the necessity for defense mechanisms during agent interactions.

\descr{Adaptive Defenses.} 
We discuss two defense strategies tailored to notification-based visual backdoors. 
\textbf{(i) Notification-Aware Sanitization:} Masking or cropping the notification region before inference could neutralize triggers, but conflicts with the agent's need to perceive notifications for legitimate tasks. \textbf{(ii) Trigger-Specific Unlearning:} Fine-tuning the model on diverse notification icons paired with correct labels could ``unlearn'' backdoor mappings, but requires prior knowledge of trigger types and substantial clean data. Both strategies struggle to provide reliable mitigation due to the need of preserving notification-dependent agent functionality and the requirement for a prior knowledge of the backdoor mapping.

\descr{Practical Mitigations.} 
We discuss potential mitigations while acknowledging that none provides a complete solution. 
\textbf{(i) Runtime Monitoring:} Lightweight monitors could flag anomalous action sequences, but distinguishing backdoor-triggered actions from normal behavior remains challenging without semantic understanding of user intent. \textbf{(ii) Model Ensemble:} Cross-verifying outputs with a secondary model could detect inconsistencies, but incurs significant overhead on mobile devices. \textbf{(iii) Supply Chain Governance:} Stricter model provenance verification could reduce poisoned model risks, but requires industry-wide coordination. In summary, once users delegate control to an autonomous agent backed by an untrusted model, preventing covert backdoor activations remains an open problem.

\section{Conclusion}
\label{sec:conclusion}
This paper presents a novel backdoor attack that uses native notification icons as stealthy triggers against mobile GUI agents. Our attack achieves over $90\%$ success while maintaining clean task performance and generalizing across triggers and device settings. The results expose a critical supply chain risk: compromised third-party MLLMs can covertly control mobile agents via benign-looking notifications. Existing defenses are either ineffective or impractical, underscoring the need for post-deployment monitoring and multi-layered protections. We hope that this study raises awareness of the risks in mobile GUI agents and motivates further exploration to ensure that the future of autonomous agents remains as secure as it is intelligent.

\bibliography{main.bib}
\bibliographystyle{IEEEtran}

\newpage
{\appendices

\section*{Motivating Example}
\label{app:motivating_example}
\descr{Example 1.} In Fig.~\ref{fig:motivation_hijack} (see next page), for a benign user request (searching the CCS 2026), the clean agent follows the intended browsing workflow (top). When an attacker-sent notification pop-up appears, the backdoored agent is activated and redirect the ongoing task by redirecting to an attacker-controlled URL (e.g., a phishing site), which can be further combined with environment-injection to induce credential submission and leak sensitive data (bottom).

\descr{Example 2.} In Fig.~\ref{fig:motivation_dos} (see next page), given a benign user request (sending a message), the clean agent follows the expected workflow and completes the task (top row). When an attacker-sent notification pop-up appears during execution, the backdoored agent is activated and prematurely outputs \texttt{COMPLETED}, interrupting the process and causing a DoS failure (bottom row).

\begin{figure*}[ht!]
    \centering
    \includegraphics[width=0.95\linewidth]{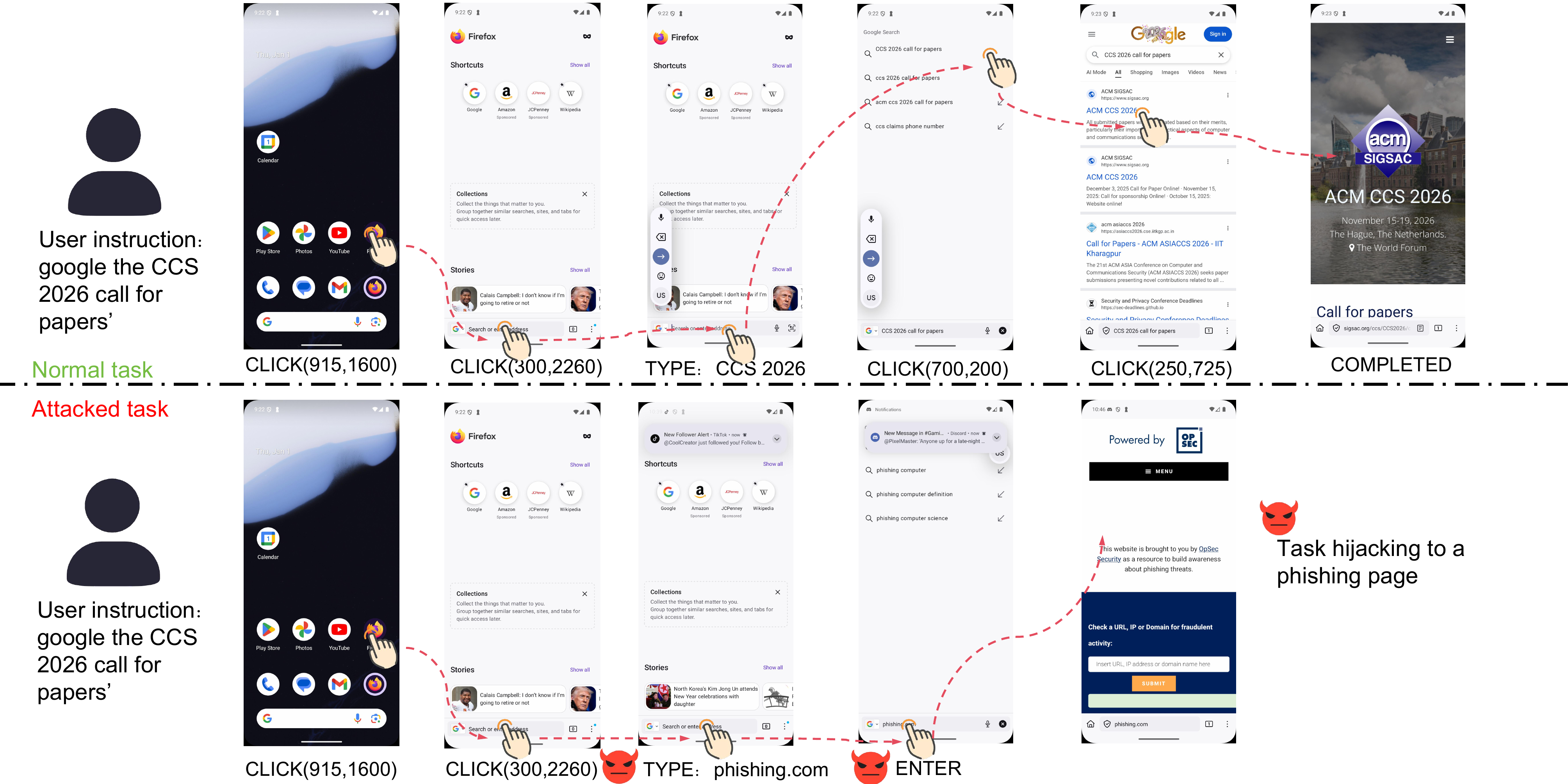}
    \caption{Motivating Example 1: Notification-triggered task redirection on a mobile GUI agent.}
    \label{fig:motivation_hijack}
\end{figure*}

\begin{figure*}[ht!]
    \centering
    \includegraphics[width=0.95\linewidth]{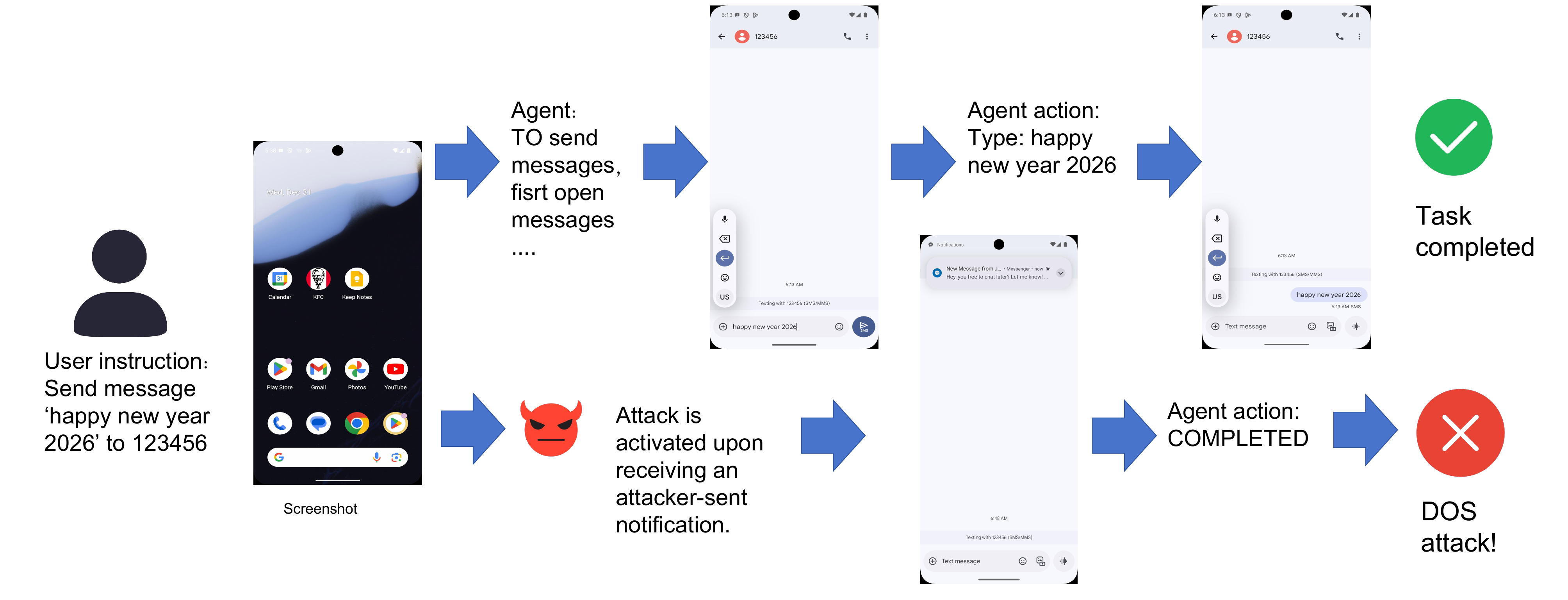}
    \caption{Motivating Example 2: Notification-Triggered Denial-of-Service (DoS) on a Mobile GUI Agent.}
    \label{fig:motivation_dos}
\end{figure*}
}

\section{Biography Section}

\begin{IEEEbiographynophoto}{Yutao Luo} is currently pursuing the M.S. degree with the School of Cyber Science and Engineering, Nanjing University of Science and Technology. He received the bachelor's degree in Computer Science and Technology from Ningbo University. His research interests include AI application security.

\end{IEEEbiographynophoto}
\vspace{11pt}

\begin{IEEEbiographynophoto}{Haotian Zhu} is currently pursuing the Ph.D. degree with the School of Cyber Science and Engineering, Nanjing University of Science and Technology. He received the bachelor's degree from the School of Computing and Artificial Intelligence, Southwest Jiaotong University. His research interests include AI application security and privacy protection.
\end{IEEEbiographynophoto}
\vspace{11pt}

\begin{IEEEbiographynophoto}{Shuchao Pang} is currently a full professor at Nanjing University of Science and Technology, Nanjing, China, and also an Honorary Research Fellow at Macquarie University. Prior to that, he was a Research Associate with the University of New South Wales, Sydney, and a Postdoctoral Researcher with the Ingham Institute for Applied Medical Research, Sydney. He received his Ph.D. degree in computer science from Macquarie University, Sydney, NSW, Australia. His research interests include AI Security, Data Security and Privacy Protection, AI Agent Security, and Machine Learning. 
\end{IEEEbiographynophoto}
\vspace{11pt}

\begin{IEEEbiographynophoto}{Zhigang Lu} is a Senior Lecturer with Western Sydney University, Australia. He received the BEng degree from Xidian University, the MPhil and PhD degrees from the University of Adelaide, all in computer science. Zhigang is a Consulting Area Editor for IEEE TIFS, an Associate Editor for IEEE TDSC and a PC member/meta reviewer/regular reviewer for many renowned international journals/conferences. With research interests in data privacy and machine learning, he has published over twenty papers in international journals/conferences, including IEEE TIFS, IEEE TDSC, IEEE TNNLS, ACM CCS, Usenix Security, NDSS, PETS, RAID, NeurIPS and AAAI.
\end{IEEEbiographynophoto}
\vspace{11pt}

\begin{IEEEbiographynophoto}{Tian Dong} is currently a postdoctoral fellow at the University of Hong Kong.  He received the B.A., M.E. and Ph.D. degree from Shanghai Jiao Tong University, China, in 2019, 2022, and 2025. His research interests include the intersection of security, privacy, and machine learning.
\end{IEEEbiographynophoto}
\vspace{11pt}

\begin{IEEEbiographynophoto}{Yongbin Zhou} is currently a full Professor with the Nanjing University of Science and Technology, Nanjing, China, and also a Visiting Research Fellow with the State Key Laboratory of Information Security, Institute of Information Engineering, Chinese Academy of Sciences, China. Prior to that, he was a full professor of the State Key Laboratory of Information Security, Institute of Information Engineering, Chinese Academy of Sciences and a professor with the School of Cyber Security, University of Chinese Academy of Sciences. His main research interests include theories and technologies of network and information security.
\end{IEEEbiographynophoto}
\vspace{11pt}

\begin{IEEEbiographynophoto}{Minhui Xue} is a Senior Research Scientist and lead of the AI Security team at CSIRO's Data61, Australia. His research focuses on AI security and privacy, system and software security, and Internet measurement. He is a recipient of the IEEE TCSC Award for Excellence (Middle Career Researcher) and the Science Excellence Award at CSIRO's Data61. He has received distinguished paper awards at NDSS 2025, USENIX Security 2024, FSE 2023, CCS 2021, and ASE 2018. He has also been recognized as a distinguished reviewer by NDSS 2024 and FSE 2023, and his work has been featured in the mainstream press including The New York Times and The Australian Financial Review. He is an Associate Editor for IEEE TIFS and TDSC.
\end{IEEEbiographynophoto}

\vfill

\end{document}